\documentclass[a4paper]{jpconf}
\usepackage[numbers,sort&compress]{natbib}
\usepackage{iopams}
\usepackage{float}
\usepackage{rotating}
\usepackage{amssymb}
\usepackage{epsfig}

\begin{document}

\title[Jet properties from $\pi^{\pm}$ - $h^{\pm}$ correlation in $p+p$ and $d$+Au collisions ......]{Jet properties from $\pi^{\pm}$ - $h^{\pm}$ correlation in $p+p$ and $d$+Au collisions at $\sqrt{s_{NN}}$ = 200 GeV}

\author{Jiangyong Jia\dag\ for the PHENIX Collaboration}

\address{\dag\ Columbia University, New York, NY 10027 and Nevis Laboratories, Irvington, NY 10533, USA}

\ead{jjia@nevis.columbia.edu}

\begin{abstract}
We review recent results on the charged pion - charged hadron
correlation in $p+p$ and $d$ + Au collisions as measured by the
PHENIX Collaboration. Properties of di-jet system, such as the jet
shape, associated hadron yield per trigger pion, and the
underlying event are extracted statistically from the
$\pi^{\pm}-h^{\pm}$ correlation function in $\Delta\phi$ and
$\Delta\eta$. For jet triggered with high $p_T$ pions ($p_T>5$
GeV/c), no apparent differences in the jet properties are seen
between $p+p$ and $d$ + Au.
\end{abstract}




\section{Introduction}

The technique of two particle correlation in relative azimuth
($\Delta\phi$) and pseudorapidity ($\Delta\eta$) is an useful tool
to access the (di-)jet properties in heavy-ion collisions.
Comparing with the traditional full jet reconstruction method, the
two particle correlation method is relatively insensitive to the
level of the underly event, thus can probe soft jets ($\lesssim$5
GeV/c); combining with event mixing technique, it can also be used
for detectors with limited acceptance.

\begin{figure}[ht]
\begin{center}
\epsfig{file=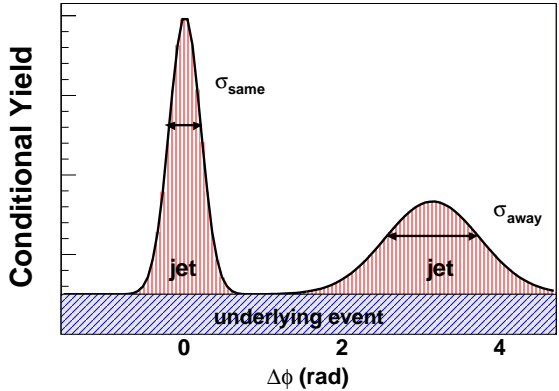,width=0.5\linewidth}
\caption{\label{fig:cartoon} Cartoon of the two particle
$\Delta\phi$ correlation. The yield of hadron per trigger
(Conditional Yield) has a di-jet part and a part corresponding to
the underlying event.}
\end{center}
\end{figure}

To leading order in QCD, high $p_T$ jets are produced back-to-back
in azimuth. This back-to-back correlation, however, is smeared by
the fragmentation process and the initial and final state
radiation, which lead to a typical di-hadron correlation function
in $\Delta\phi$ as shown schematically in
Figure.\ref{fig:cartoon}. The associated hadron yield per trigger
$\pi^{\pm}$ (conditional yield or CY) can be parameterized by a
constant plus a double gauss function,
\begin{equation}\label{eq:cy1}
\frac{1}{N^0_{\rm{trig}}}\frac{dN_0}{d\Delta\phi} =
B+\frac{\rm{N}_{S}}{\sqrt{2\pi}\sigma_{N}}
e^{\frac{-\Delta\phi^2}{2\sigma_{N}^2}} + \frac{\rm{N}_{A}}
{\sqrt{2\pi}\sigma_{F}}
e^{\frac{-(\Delta\phi-\pi)^2}{2\sigma_{F}^2}},
\end{equation}
In this analysis, everything about the (di-)jet is extracted from
this parameterization. The peaks in the same side ($\Delta\phi=0$)
and the away side ($\Delta\phi=\pi$) represent the intra-jet and
di-jet correlation, respectively. The widths of the peaks are
controlled by the jet fragmentation momentum $j_T$ and the parton
transverse momentum $k_T$~\cite{Rak:2004gk,Jia:2004sw}:
$\sigma_{same}\propto j_{Ty}$, $\sqrt{\sigma^2_{away}-
\sigma^2_{same}}\propto k_{Ty}$, where the subscript ``$y$''
represent the 1D projection in transverse plane; The integrals of
the peaks, $N_{S}$ and $N_A$ give the total number of hadrons
associated with the trigger hadrons in the same side and the away
side; The pedestal level beneath the jet structure, $B$,
represents contributions from the underlying event.

The presence of the medium can modify the di-hadron correlation.
Recent results from RHIC~\cite{Adler:2002tq,dan,wolf} and
SPS~\cite{Agakichiev:2003gg} indicate a complicated modification
of the jet structure in both the same side and the away side. The
same side jet become elongated in $\Delta\eta$ but is relatively
unmodified in $\Delta\phi$ direction. Meanwhile, the away side
correlation become much broader and suppressed, which indicates
strong interaction of the jets with the dense medium. In order to
achieve quantitative understandings of the modifications of jets
in the medium, one has to obtain an accurate baseline measurements
of jet correlation in $p+p$ and $p+A$ collisions. Di-hadron
correlation in $p+p$ collisions probes basic QCD effects such as
jet fragmentation, initial and final state radiations; while
correlation in $p+A$ collisions gives handle on the various
initial state effects such as shadowing and jet broadening in cold
nuclear medium.

We focus the physics discussions on three aspects of the
$\pi^{\pm}-h^{\pm}$ correlations in Figure.\ref{fig:cartoon}: jet
shape, jet yield and the underlying event. But before doing that,
we briefly discuss the identification of high $p_T$ charged pion
and the techniques used to extract the jet properties.

\section{Analysis}
\label{sec:ana}
\subsection{$\pi^{\pm}$ identification}

PHENIX identifies high momentum charged pions with the RICH and
EMCal detectors. Charged particles with velocities above the
Cherenkov threshold of $\gamma_{\rm{th}} = 35$ (CO$_{2}$ radiator)
emit Cherenkov photons, which are detected by photo-multiplier
tubes (PMTs) in the RICH~\cite{Aizawa:2003zq}. This threshold
corresponds to 18 MeV/$c$ for electrons, 3.5 GeV/$c$ for muons and
4.9 GeV/$c$ for charged pions. In a previous PHENIX
publication~\cite{Adler:2003au}, we have shown that charged
particles with reconstructed $p_T$ above 4.9 GeV/$c$, which have
an associated hit in the RICH, are dominantly charged pions and
background electrons from photon conversions. The efficiency for
detecting charged pions rises quickly past 4.9 GeV/$c$, reaching
an efficiency of $>90\%$ at $p_T>6$ GeV/$c$.

\begin{figure}[ht]
\begin{center}
\epsfig{file=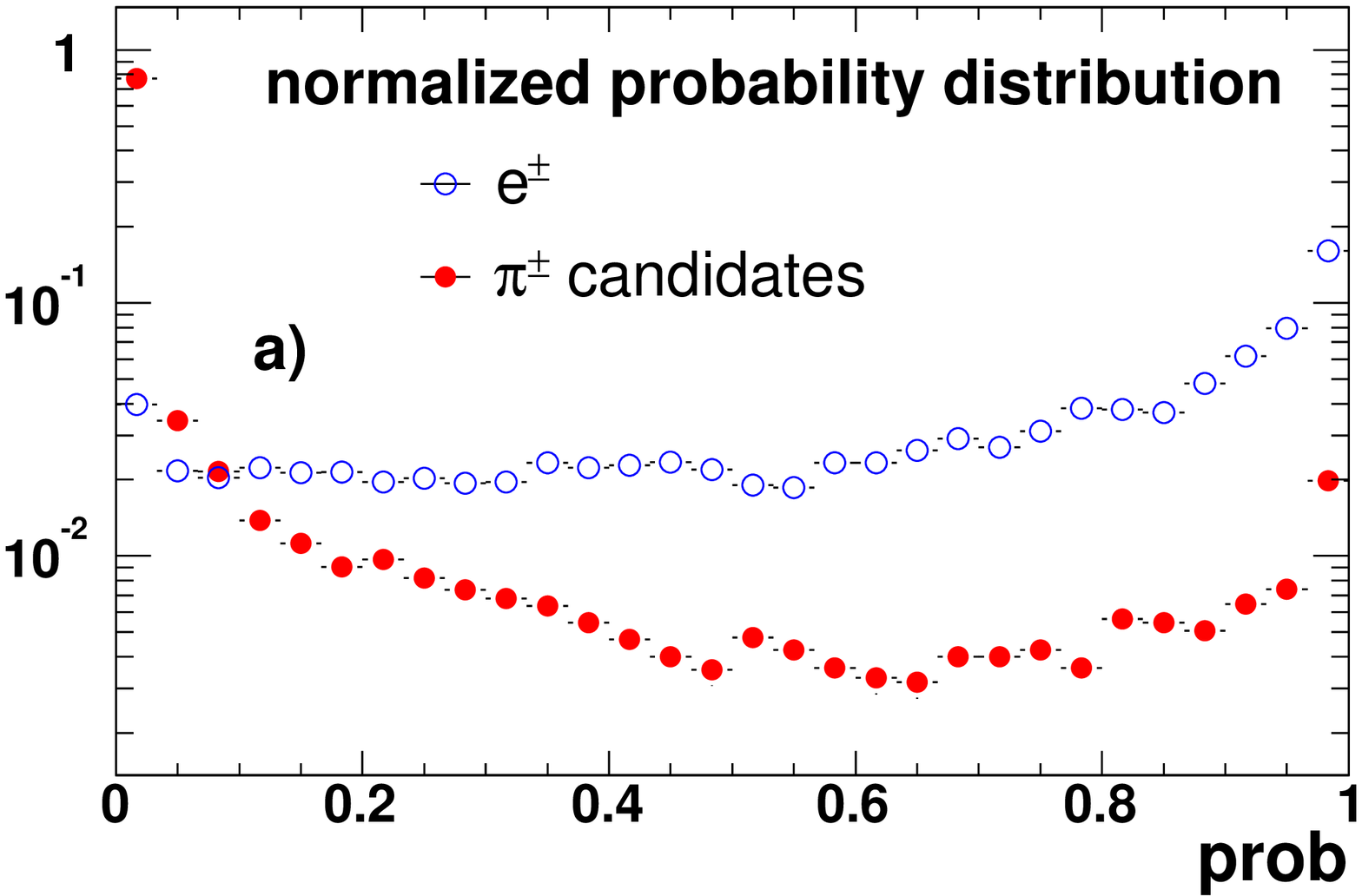,width=0.45\linewidth}
\epsfig{file=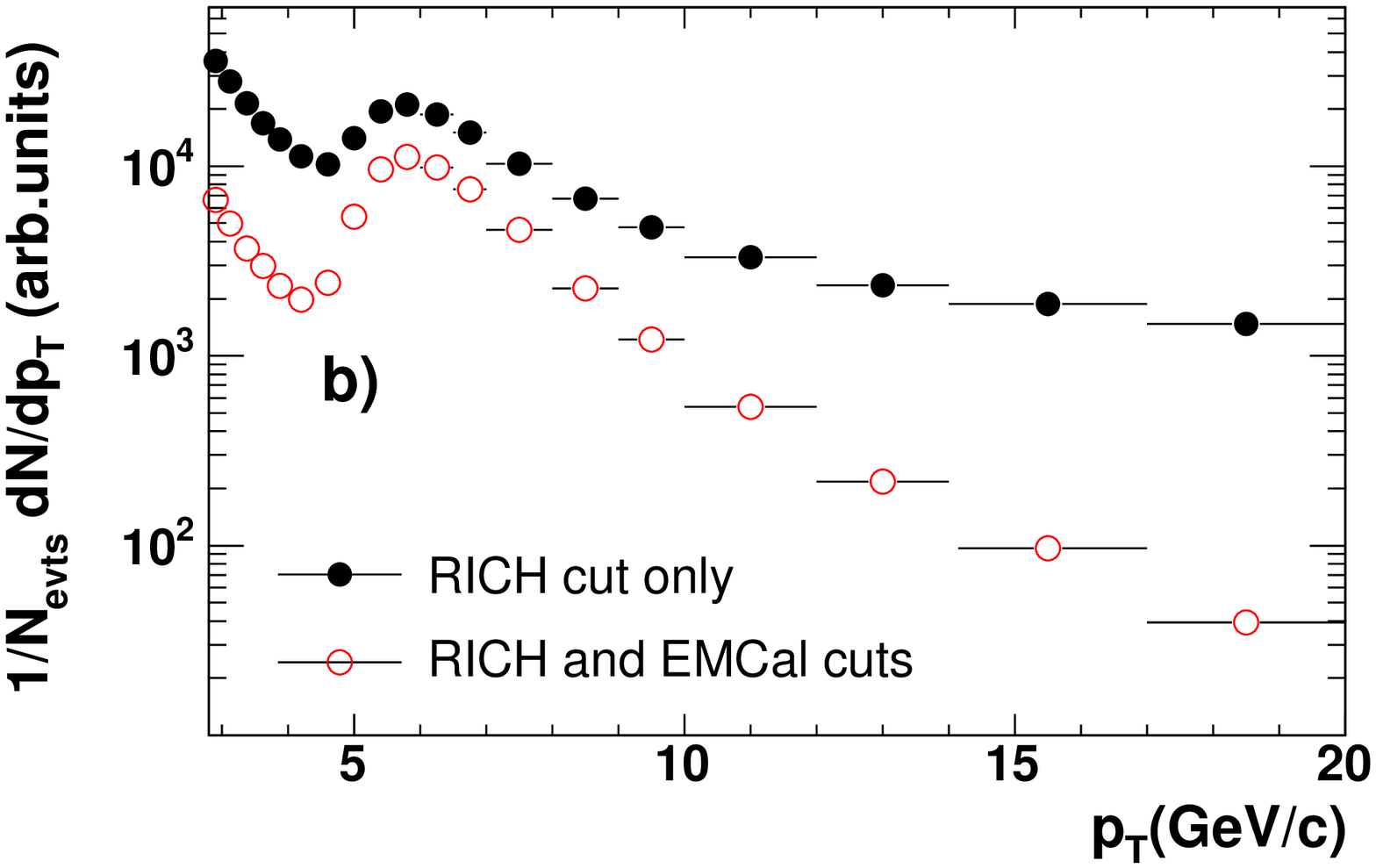,width=0.47\linewidth}
\caption{\label{fig:prob} a) The probability distribution for
charged pion candidates (solid circles) and electrons (open
circles), where the integrals of the histograms have been
normalized to one. b) Raw pion spectra with matching hit at RICH
only (solid circles) and with both RICH and EMCal cuts (open
circles).}
\end{center}
\end{figure}

To reject the conversion backgrounds in the pion candidates, the
shower information at the EMC is used. Since most of the
background electrons are genuine low $p_T$, they can be rejected
by requiring a large shower energy in the EMC. In this analysis, a
momentum dependent energy cut at EMC is applied: $E>0.3+0.15 p_T$.
Additional electron rejection comes from the $\chi^2$ variable,
\begin{eqnarray}
\chi^2
=\sum_{i}\frac{(E^{\rm{meas}}_{i}-E^{\rm{pred}}_{i})^2}{\sigma_{i}^2}
\label{eq:chisq}
\end{eqnarray}
where $E^{\rm{meas}}_{i}$ is the energy measured at tower $i$,
$E^{\rm{pred}}_{i}$ is the predicted energy for an electromagnetic
particle of total energy $\sum_{i}E^{\rm{meas}}_{i}$ and
$\sigma_{i}$ is the standard deviation for $E^{\rm{pred}}_{i}$.
Both $E^{\rm{pred}}_{i}$ and $\sigma_{i}$ are obtained from the
electron test beam data. EM shower is more compact than hadronic
shower, thus has a smaller $\chi^2$ value. The $\chi^2$ value is
then mapped to the probability ($prob$) for a shower being an EM
shower. $prob$ ranges from 0 to 1, with a flat distribution
expected for EM showers and a distribution peaked around 0 for
hadronic showers. Figure.~\ref{fig:prob}a shows the normalized
$prob$ distribution for the pion candidates and electrons. A cut
of $prob$~$< 0.2$ selects pions with an efficiency of $\gtrsim
80$\%. Since we are interested in per-triggered yield, the
detailed knowledge of the pion efficiency is not necessary. The
raw pion spectra for requiring only RICH cut and both RICH and
EMCal cuts are shown in Figure.~\ref{fig:prob}b, the difference
between the two is mostly due to electron background. The sample
of charged pion used in the correlation analysis is from 5 to 16
GeV/c, with an purity better than 95\%.

\subsection{Extracting jet properties}
The correlation functions are generally defined as
\begin{eqnarray}
C\left(\Delta\phi\right) = \frac{N_{\rm{cor}}\left( \Delta\phi
\right)} {N_{\rm{mix}}\left( \Delta\phi \right)} \textrm{ in
azimuth and  } C\left(\Delta\eta\right) = \frac{N_{\rm{cor}}\left(
\Delta\eta \right)} {N_{\rm{mix}}\left( \Delta\eta \right)}
\end{eqnarray}
in pseudorapidity. $N_{\rm{cor}}$ and $N_{\rm{mix}}$ represent the
same-event pair distribution and mixed-event pair distributions,
respectively. The mixing is done by pairing trigger $\pi^{\pm}$
with charged hadrons from events having similar collision vertex
and centrality as the $\pi^{\pm}$. Ref.\cite{Jia:2004sw} has shown
that the correlation function and the conditional yield are
related to each other by just a constant,
\begin{eqnarray}\label{eq:cy2}
\frac{1}{N^0_{\rm{trig}}}\frac{dN_0}{d\Delta\phi}
=\frac{R_{\Delta\eta}}{N_{\rm{trig}}\epsilon}\frac{N_{\rm{cor}}\left(\Delta\phi\right)}{\frac{2\pi
N_{\rm{mix}}\left(\Delta\phi\right)}{\int d\Delta\phi
N_{\rm{mix}}\left(\Delta\phi\right)}}
\end{eqnarray}
where $N^0_{\rm{trig}}$ and $N_{\rm{trig}}$ are the true and
detected number of triggers respectively, and $\epsilon$ is the
average single particle efficiency for the associated particles in
$2\pi$ in azimuth and $\pm0.35$ in pseudo-rapidity. $R_{\Delta
\eta}$ accounts for the loss of jet pairs outside PHENIX pair
acceptance of $|\Delta \eta|<0.7$.

The two gauss functions in Eq.\ref{eq:cy1} describe the
$\Delta\phi$ distribution of the jet signal. The jet signal can
also be presented in any other pair variables, such as
$\Delta\eta$, trigger $p_T$ ($p_{T,\rm{trig}}$), associated hadron
$p_T$ ( $p_{T,\rm{assoc}}$), $p_{\rm{out}} =
p_{T,assoc}\sin\Delta\phi$, $x_E =
\frac{p_{T,assoc}\cos\Delta\phi}{p_{T,trig}}$, di-hadron mass and
di-hadron $p_T$. For every pair variable, we use a statistical
weighting method to account for the acceptance correction.
According to Eq.\ref{eq:cy2}, each pair on average needs a
$\Delta\phi$ dependent correction factor, $w(\Delta\phi)$,
\begin{eqnarray}
\label{eqn:weight} w(\Delta\phi) =
\frac{R_{\Delta\eta}}{N_{\rm{trig}}\epsilon}\frac{1}{\frac{2\pi
N_{\rm{mix}}\left(\Delta\phi\right)}{\int d\Delta\phi
N_{\rm{mix}}\left(\Delta\phi\right)}}
\end{eqnarray}
When this factor is used as the weight in filling the $x_E$
histograms for both real and mixed pairs, we obtain
\begin{eqnarray}
\frac{1}{N^0_{\rm{trig}}}\frac{dN_0}{dx_E} = \sum_{\rm{real}}
\delta(x_E) w(\Delta\phi)
\end{eqnarray}
for the same-event pair distribution. Thus according to
Eq.\ref{eq:cy1}, $x_E$ distribution for jet pairs equals to
\begin{eqnarray}
\label{eq:statdndxe}
\frac{1}{N^0_{\rm{trig}}}\frac{dN_0^{\rm{jet}}}{dx_E} =
\sum_{\rm{real}} \delta(x_E) w(\Delta\phi) - C \sum_{\rm{mix}}
\delta(x_E) w(\Delta\phi)\quad.
\end{eqnarray}
where
\begin{eqnarray}
C =\frac{BR_{\Delta\eta}}{N_{\rm{trig}}\epsilon}\frac{2\pi}{\int
d\Delta\phi N_{\rm{mix}}\left(\Delta\phi\right)} \quad.
\end{eqnarray}

Replacing $x_E$ with any pair variables, we obtain other jet pair
distributions. However, the integral of the jet yield should be
conserved independent of the pair variable used, {\it i.e.}:
\begin{eqnarray}
\int d\Delta\phi\frac{dN_0^{\rm{jet}}}{d\Delta\phi}= \int
dx_E\frac{dN_0^{\rm{jet}}}{dx_E}=\int dp_{T,\rm{assoc}}
\frac{dN_0^{\rm{jet}}}{dp_{T,\rm{assoc}}} = \int
dp_{\rm{out}}\frac{dN_0^{\rm{jet}}}{dp_{\rm{out}}}
\end{eqnarray}

\section{Results}
\label{sec:res}
\subsection{Jet shape}
\begin{figure}[ht]
\begin{center}
\begin{tabular}{c}
\resizebox{0.8\columnwidth}{!}{\includegraphics{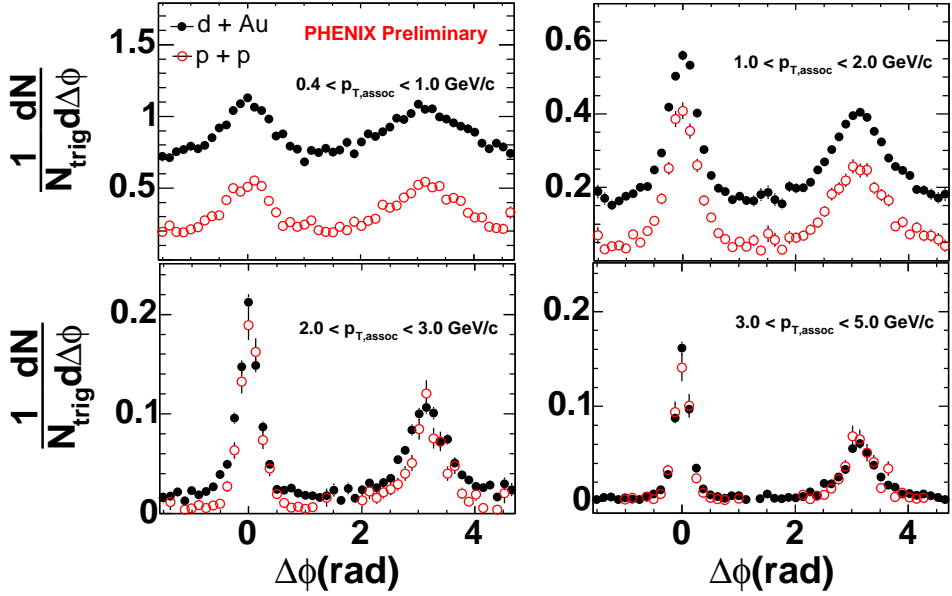}}
\end{tabular}
\caption{\label{fig:dphi} Corrected conditional pair distributions
for $p+p$ and minimum bias $d$+Au collisions. The trigger
$\pi^{\pm}$ are correlated with hadrons with $p_{T,\rm{assoc}}$
$0.4-1.0$ GeV/$c$, $1.0-2.0$ GeV/$c$, $2.0-3.0$ GeV/$c$ and
$3.0-5.0$ GeV/$c$ (from top to bottom and left to right).}
\end{center}
\end{figure}

In the following discussion, the trigger $\pi^{\pm}$ $p_T$ is
always from 5 to 10 GeV/$c$, unless specified otherwise.
Figure.\ref{fig:dphi} shows the $\pi^{\pm} -h^{\pm}$ $\Delta\phi$
distribution from $p+p$ and $d$ + Au collisions for several range
of $p_{T,assoc}$. The widths decrease with increasing
$p_{T,assoc}$, which is consistent with narrowing of the jet cone
for larger $p_{T,assoc}$. It is interesting to notice that a large
fraction of all hadrons in the event are associated with the
trigger, thus are originated from the hard-scattered partons. Even
for $p_{T,assoc}$ as low as $0.4-1$ GeV/$c$, about 51\% hadron
yield in $p+p$ (27\% in $d$ + Au) comes from the jet
fragmentation.

\begin{figure}[ht]
\begin{center}
\epsfig{file=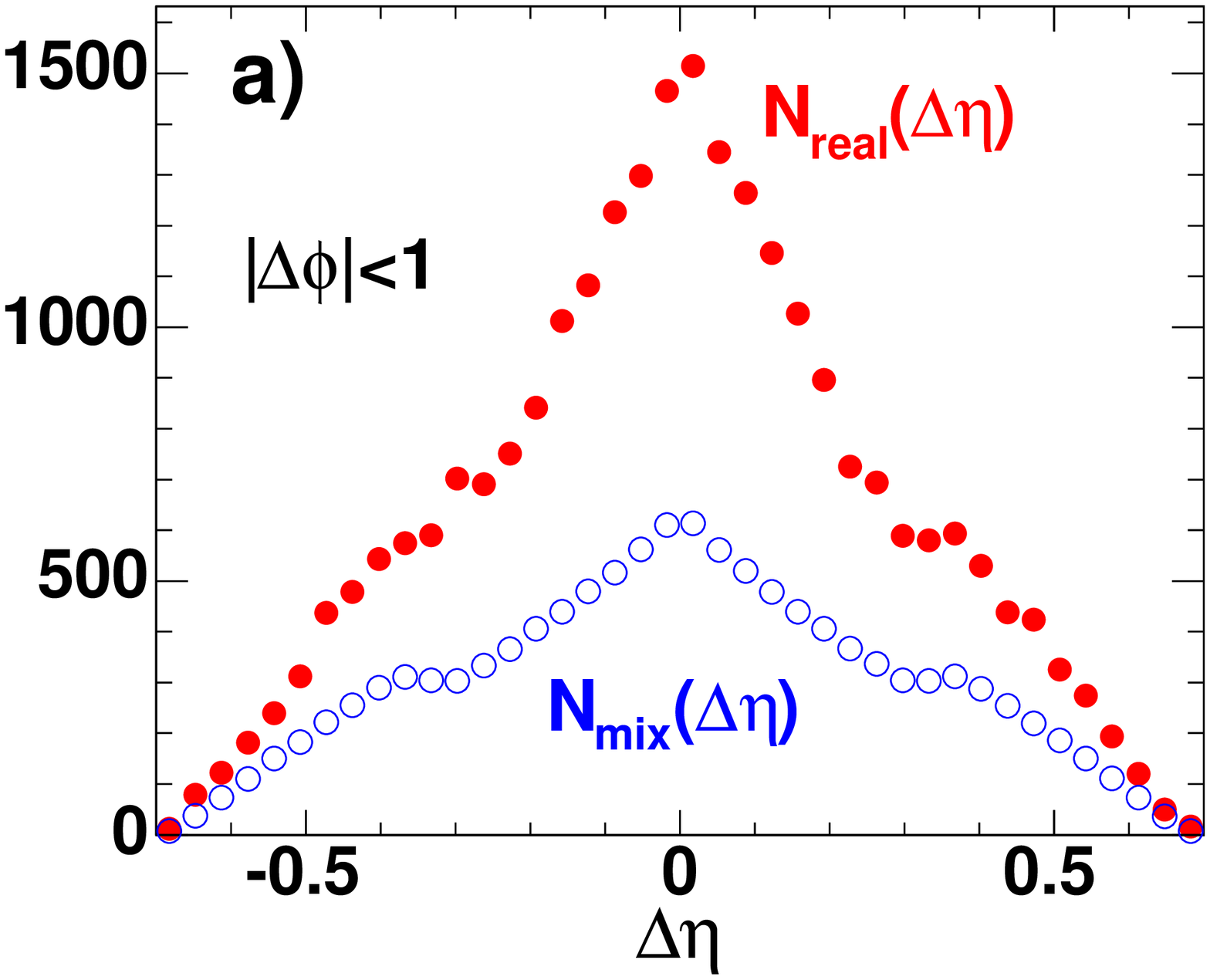,width=0.4\linewidth}\epsfig{file=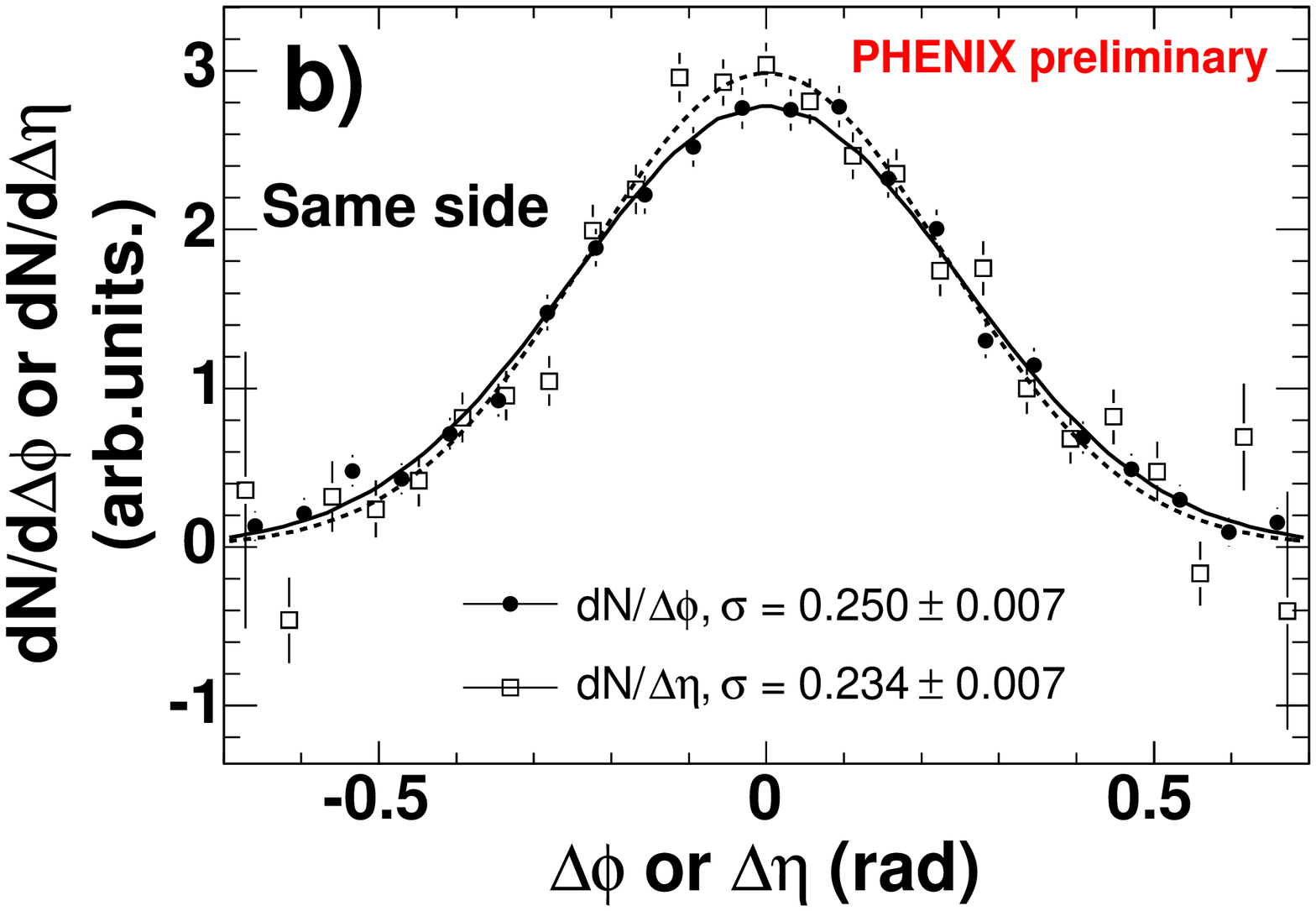,width=0.45\linewidth}
\caption{\label{fig:eta1} a) The same-event and mixed-event pair
distribution in $\Delta\eta$, b) the correlation function in
$\Delta\eta$ (open boxes) and $\Delta\phi$ (filled circles).}
\end{center}
\end{figure}

\begin{figure}[ht]
\begin{center}
\epsfig{file=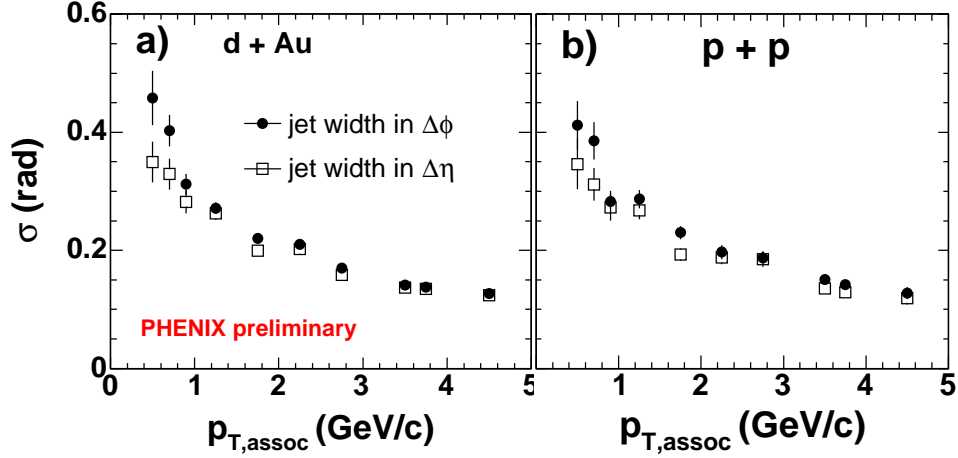,width=0.8\linewidth}
\caption{\label{fig:eta2}The comparison of jet width as function
of $p_{T,\rm{assoc}}$ in $\Delta\phi$ (solid circles) and
$\Delta\eta$ (open boxes) from $\pi^{\pm}-h^{\pm}$ correlation. a)
results for $d$ + Au. b) results for $p+p$.}
\end{center}
\end{figure}

Using the event mixing technique, we also measure the jet shape in
$\eta$. For the single acceptance of $|\eta|<0.35$, the pair
acceptance in pseudorapidity is limited to be $|\Delta\eta|<0.7$.
Figure.\ref{fig:eta1}a shows the same event and mixed event
$\Delta\eta$ distribution for $1<p_{T,assoc}<2$ GeV/$c$, where a
cut of $|\Delta\phi|<1$ is used to select only same side jet
pairs. The mixed-event pair distribution is not a perfect triangle
due to a gap around $\eta=0$ in PHENIX central arm detectors. The
ratio of the two distributions gives the jet shape in
$\Delta\eta$. It is shown in Figure.\ref{fig:eta1}b and compared
with the jet shape in $\Delta\phi$. There is no significant
difference between the two and the extracted widths are consistent
in both directions. We extend this comparison to other associated
hadron $p_T$ ranges and summarize the results in
Figure.\ref{fig:eta2}. The overall agreement between the jet
widths in $\Delta\eta$ and $\Delta\phi$ is pretty good, except at
small $p_{T,\rm{assoc}}$, where the width in $\Delta\eta$ is
systematically lower than that in $\Delta\phi$. The fact that this
discrepancy exist in both $p+p$ and $d$ + Au collisions indicates
that this deviation is likely due to the systematics of the
fitting in a limited $\Delta\eta$ range rather than any real
physics effect in $d$ + Au.

\begin{figure}[ht]
\begin{tabular}{cc}
\begin{minipage}{0.5\linewidth}
\begin{flushleft}
\epsfig{file=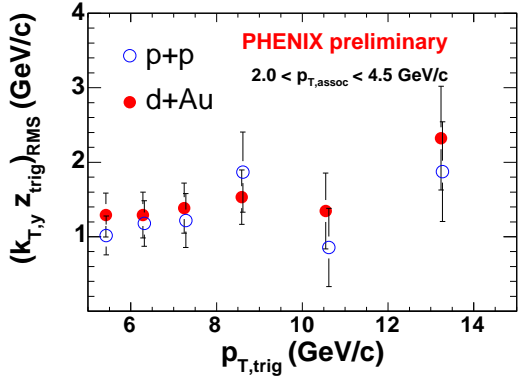,width=0.9\linewidth}
\end{flushleft}
\end{minipage}
\begin{minipage}{0.5\linewidth}
\begin{flushleft}
\caption{\label{fig:zkt} The comparison of
$(z_{\rm{trig}}k_{T_y})_{\rm{RMS}}$ values between $d$ + Au
(filled circles) and $p+p$ (open circles) as function of
$p_{T,trig}$.}
\end{flushleft}
\end{minipage}
\end{tabular}
\end{figure}

From the measured jet width, one can calculate the rms value of
$zk_{Ty}$~\cite{Jia:2004sw}, $(zk_{Ty})_{RMS} =
\sqrt{\left<z^2k^2_{Ty}\right>}$, for both $p+p$ and $d$ + Au. The
resulting $(zk_{Ty})_{RMS}$ is plotted as function of trigger
$p_T$ in Figure.\ref{fig:zkt}. It looks quite similar between
$p+p$ and $d$ + Au. The difference of $(zk_{Ty})_{RMS}$, averaged
over $p_T$, is $\langle z^2k_{Ty}^{2} \rangle_{\rm{dAu}} - \langle
z^2k_{Ty}^{2} \rangle_{\rm{pp}} = 0.64\pm 0.78\pm 0.42$
(GeV/$c$)$^{2}$, which is consistent with 0. According to various
theoretical estimations~\cite{Wang:1998ww}, the typical
contribution to $\langle k_{Ty}^{2} \rangle$ from multiple
scattering is 1 (GeV/$c$)$^2$ in central $d$ + Au collisions,
while the contribution from initial and final radiation is much
larger (around 8 (GeV/$c$)$^2$)~\cite{Boer:2003tx}. The small
multiple scattering contribution might have been washed out by the
much larger contributions from initial and final radiation, which
explains the lack of difference between the two systems.

\begin{figure}[ht]
\begin{center}
\epsfig{file=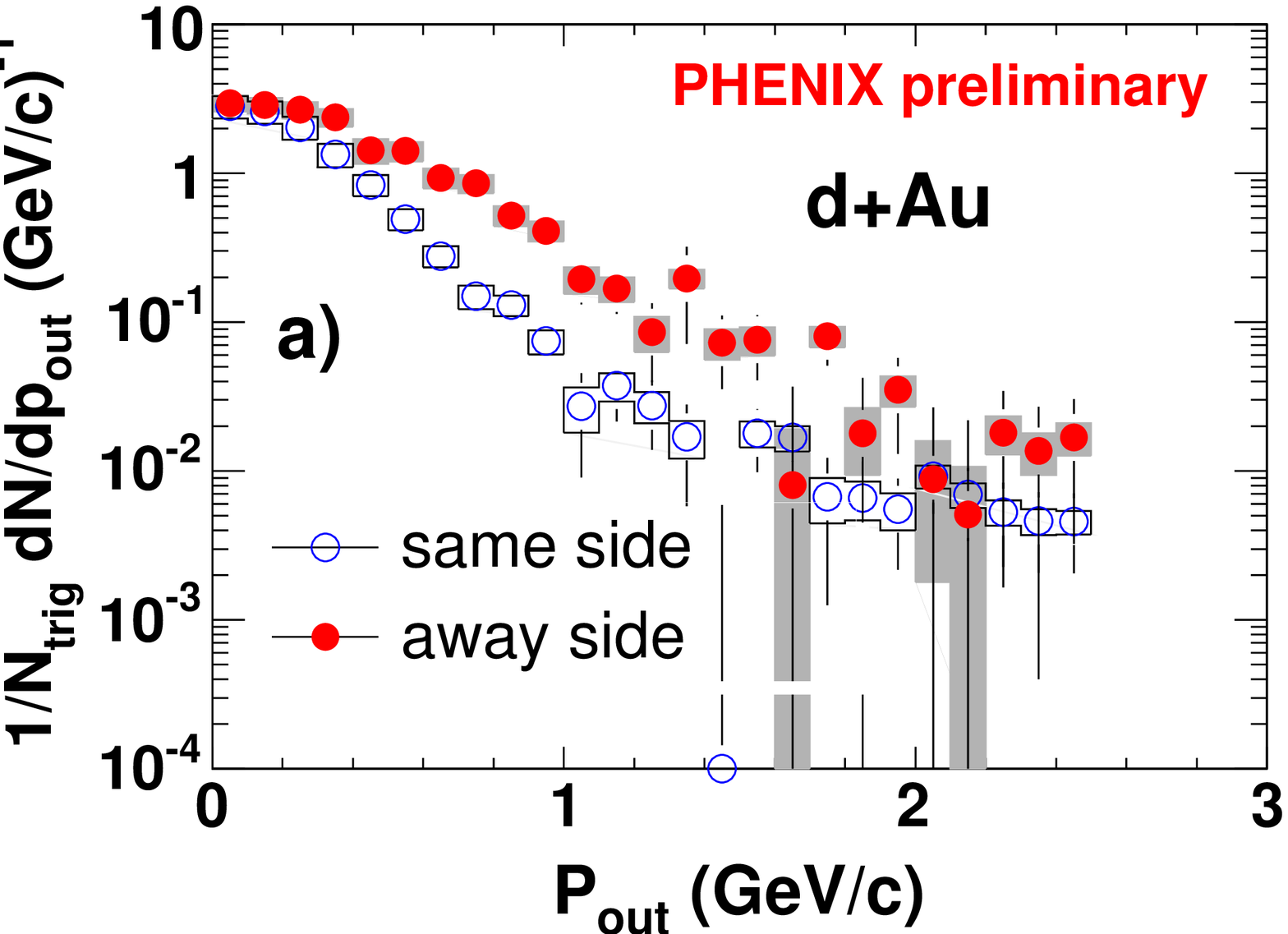,width=0.45\linewidth}
\epsfig{file=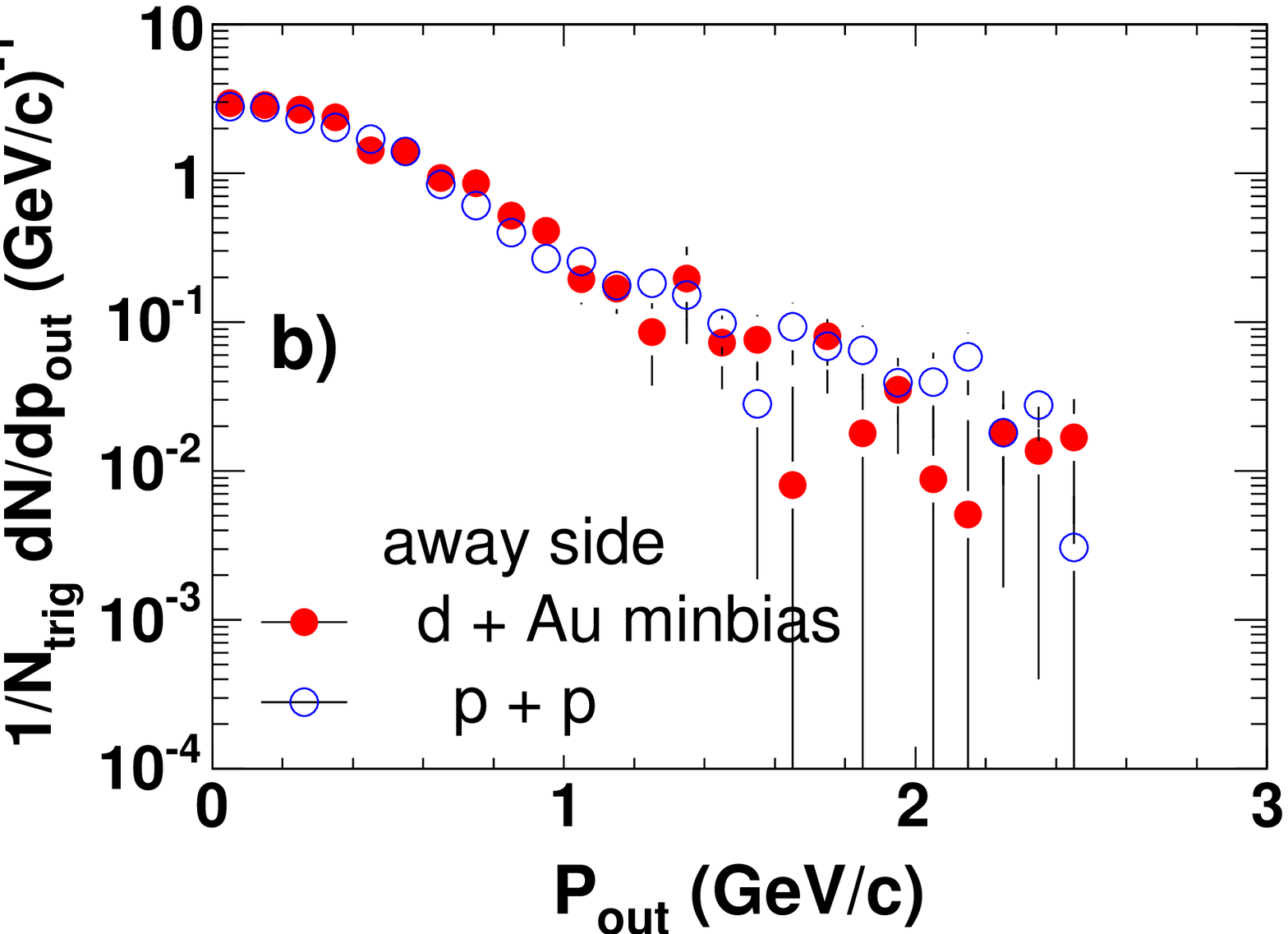,width=0.45\linewidth}
\caption{\label{fig:daupppout} a) Same side and away side
$p_{\rm{out}}$ distributions in $d$ + Au collisions. b)  The away
side $p_{\rm{out}}$ distributions compared between $p+p$ and $d$ +
Au.}
\end{center}
\end{figure}

An alternative but more direct way in studying the away side $k_T$
broadening is through the $p_{out}$ distribution. For small
angles, $p_{out}$ has simple relation to $j_T$ and
$k_T$\cite{Jia:2004sw}:
\[
\langle {p_{out,same}^2 }\rangle  \approx \langle {j_{T,y}^2 }
\rangle + x_E^2 \langle {j_{T,y}^2 }\rangle \]\[ \langle
{p_{out,away}^2 } \rangle \approx \langle {j_{T,y}^2 } \rangle +
x_E^2 \langle {j_{T,y}^2 } \rangle + 2x_E^2 \langle {z^2k_{T,y}^2
} \rangle
\]
Thus the difference of same side and away side $p_{out}$
distributions directly reflects the contribution from $k_T$.
\begin{eqnarray}
2x_E^2  \left\langle {z^2k_{T,y}^2 } \right\rangle  \approx
\left\langle {p_{out,same}^2 } \right\rangle  - \left\langle
{p_{out,away}^2 } \right\rangle
\end{eqnarray}
Figure.~\ref{fig:daupppout}a shows the same side and away side
$p_{out}$ distribution. There is a significant difference between
the two, which reflects the contribution from $k_T$. Both
distributions have a gauss shape at small $p_{out}$ followed by an
excess at large $p_{out}$. The gauss part presumably is due to the
jet fragmentation (in both the same and away side) and intrinsic
$k_T$ (away side only), while the excess is evidence for hard
radiation contribution of the outgoing partons. Since the away
side $p_{out}$ carries information about $k_T$, we compare between
$p+p$ and $d$ + Au to see whether there is hint of additional
$k_T$ broadening in $d$ + Au. The comparisons of away side
$p_{out}$ distributions are shown in Figure.\ref{fig:daupppout}b,
no apparent differences are observed, consistent with the
observations that $(zk_{T_y})_{\rm{RMS}}$ are similar between
$p+p$ and $d$ + Au.

\subsection{Jet yields}

\begin{figure}[ht]
\begin{center}
\epsfig{file=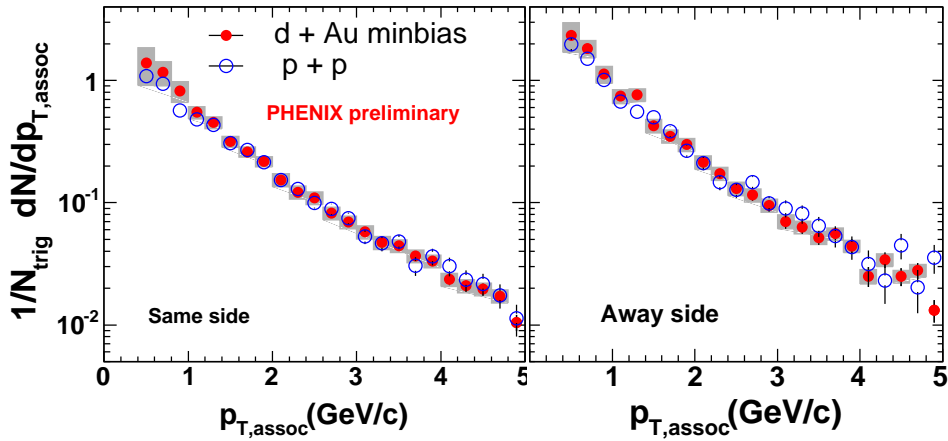,width=0.8\linewidth}
\caption{\label{fig:daupp1} Jet pair distribution as function of
$p_{T,assoc}$ for same side (right panel) and away side (left
panel) in $p+p$ and $d$ + Au.}
\end{center}
\end{figure}

The same side and away side $p_T$ distributions of the charged
hadrons associated with jets are plotted in Fig\ref{fig:daupp1},
comparing between $p+p$ and $d$ + Au collisions. The same side
yield is related to the di-hadron fragmentation, since both
particles comes from the same jet, while the away side yield
depends on two independent fragmentation functions: one parton
fragments to produce the trigger, while the second parton produces
the associated hadron. No apparent differences are seen between
$p+p$ and $d$ + Au; this observation is in contradiction to some
recombination model prediction~\cite{Hwa:2004sw}, in which a
significant difference is expected due to shower-thermal
contribution.

\begin{figure}[ht]
\begin{center}
\epsfig{file=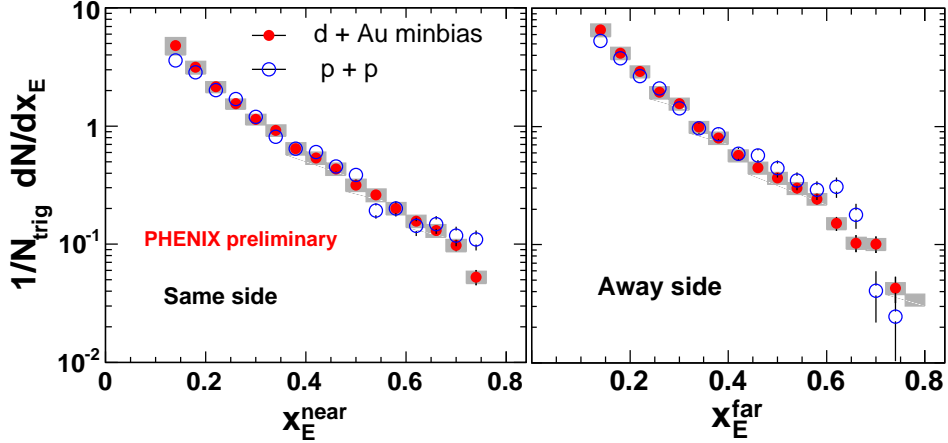,width=0.8\linewidth}
\caption{\label{fig:daupp2} $x_E$ distributions as function of
$p_{T,assoc}$ for same side (right panel) and away side (left
panel) in $p+p$ and $d$ + Au.}
\end{center}
\end{figure}

Di-hadron correlation also gives the $x_E$ distribution
$\frac{1}{{N_{trig}}}\frac{{dN_h}}{{dx_E }}$, where $x_E =
z_{\rm{assoc}}/z_{\rm{trig}}$. When di-jet $p_T$ imbalance is
ignored and $z_{\rm{assoc}}\ll z_{\rm{trig}}$, $z_{trig}$ varies
very slowly with $z_{\rm{assoc}}$. Hence the $x_E$ distribution is
closely related to the fragmentation function $D(z)$,
\begin{equation}
\frac{1}{{N_{trig}}}\frac{{dN_h}}{{dx_E }} \approx z_{trig}
D(z)\quad.
\end{equation}
Figure.\ref{fig:daupp2} shows the measured $x_E$ distribution
between $p+p$ and $d$ + Au. Again, no difference is seen between
the two collision systems in both the same side and away side.

\begin{figure}[ht]
\begin{center}
\epsfig{file=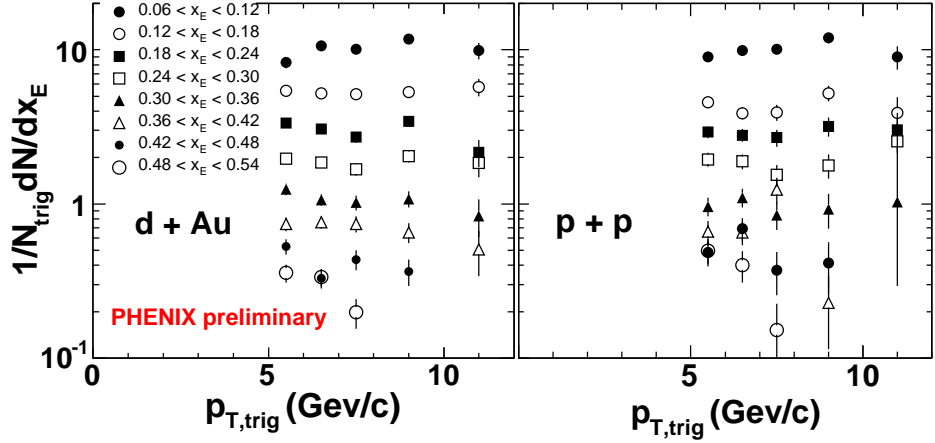,width=0.8\linewidth}
\caption{\label{fig:daupp3} Away side conditional yield as
function of $p_{T,\rm{trig}}$ for different ranges of $x_E$ in a)
minimum-bias $d$ + Au collisions and b) $p+p$ collisions.}
\end{center}
\end{figure}

In $e^+e^-$ or $p+p$ collisions, the fragmentation functions
$D(z)$ are known to approximately scale, {\it i.e.} are
independent of jet energy. To check whether this is still true in
$d$ + Au collisions, we plot in Figure.~\ref{fig:daupp3} the
conditional yields as function of trigger $p_T$ in different
ranges of $x_E$ for both $p+p$ and $d$ + Au. The amount of
variation is within $\pm$ 25\% for $p_T$ from $5-12$ GeV/$c$ with
very little difference between the two systems. So we conclude
that the evolution of the jet fragmentation as function of jet
energy is very similar between $p+p$ and $d$ + Au.

\subsection{Comments on underlying event}
Events triggered by high $p_T$ hadrons not only contain particles
originated from the two hard-scattered partons, but also those
come from soft multiple interaction and the beam remnants.
Underlying event in $p+p$ and $d$ + Au collisions refers to all
hadrons except those from the two outgoing hard-scattered partons,
which includes contributions from the beam remnants and initial
and final state radiation~\cite{Affolder:2001xt}. The physics of
the underlying event is poorly known due to its non-perturbative
nature. It is often studied phenomenologically with various QCD
Monte-carlo models that have been tuned to fit the
data~\cite{Sjostrand:2004pf}. Underlying event has been studied
extensively at the Tevatron
energy~\cite{Acosta:2004wq,Affolder:2001xt}. Similar studies at
the RHIC are very useful in understanding it's dependence on
$\sqrt{s}$, and can provide valuable constrains on the underlying
event physics at the LHC.

\begin{figure}[ht]
\begin{center}
\begin{tabular}{c}
\resizebox{0.9\columnwidth}{!}{\includegraphics{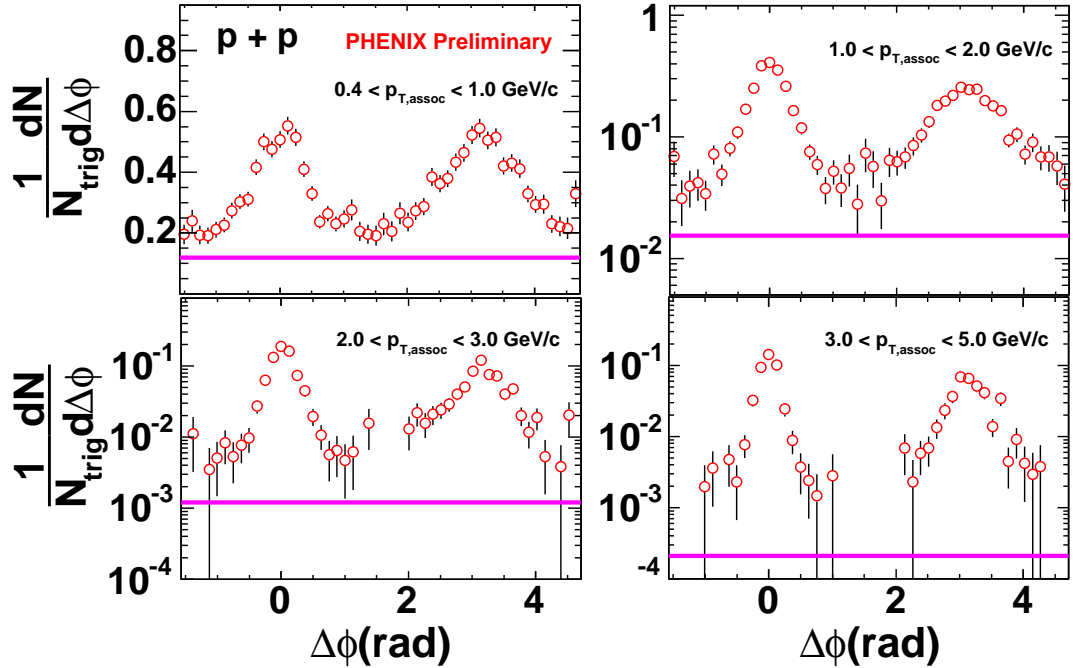}}
\end{tabular}
\caption{\label{fig:dphipp} Corrected condition yield in
$\Delta\phi$ for $p+p$ collisions (from Figure.\ref{fig:dphi}).
The thick solid line represents the average level for minimum bias
events, i.e. it is equal to Yield$_{\rm{pp}}/(2\pi)$.}
\end{center}
\end{figure}

Figure.\ref{fig:dphipp} shows the jet pair distribution in $p+p$
collisions, reproduced from Figure.\ref{fig:dphi}. The pedestal in
the $\Delta\phi$ correlation, which represents the underlying
event contribution, decreases quickly and becomes negligible at
$p_{T,assoc}>2$ GeV/$c$. However, the level corresponding to
minimum bias $p+p$ events, denoted by the thick horizontal line,
seems to decrease even faster. Since minimum bias event has small
hard-scattering contribution, the relative abundance of the
pedestal in triggered events over the minimum bias events
indicates that most of the underlying event comes from the initial
or final state radiation of the hard-scattered partons.

\begin{figure}[ht]
\begin{center}
\epsfig{file=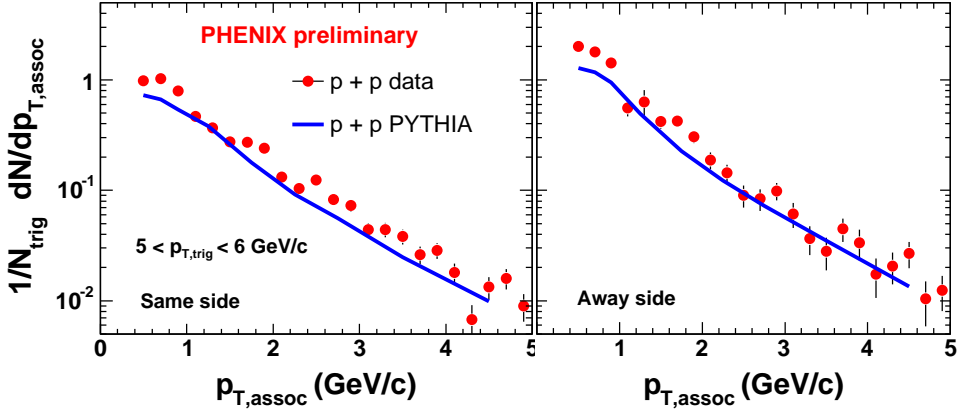,width=0.8\linewidth}
\caption{\label{fig:un1} Conditional yield with trigger pion in
$5<p_{T,trig}<6$ GeV/$c$, compared with PYTHIA 6.131 for both same
side (left panel) and away side (right panel).}
\end{center}
\end{figure}

\begin{figure}[ht]
\begin{center}
\epsfig{file=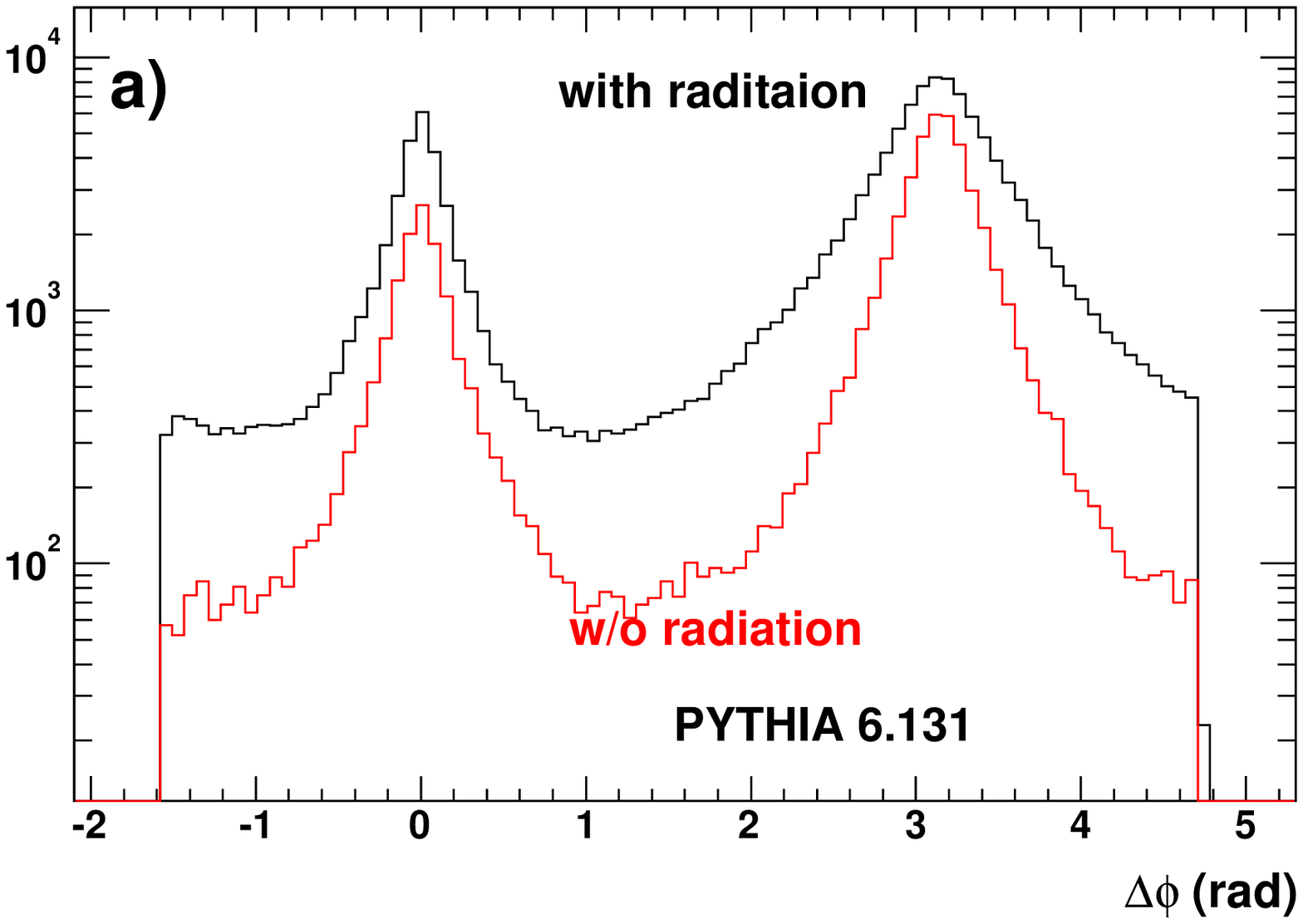,width=0.47\linewidth}
\epsfig{file=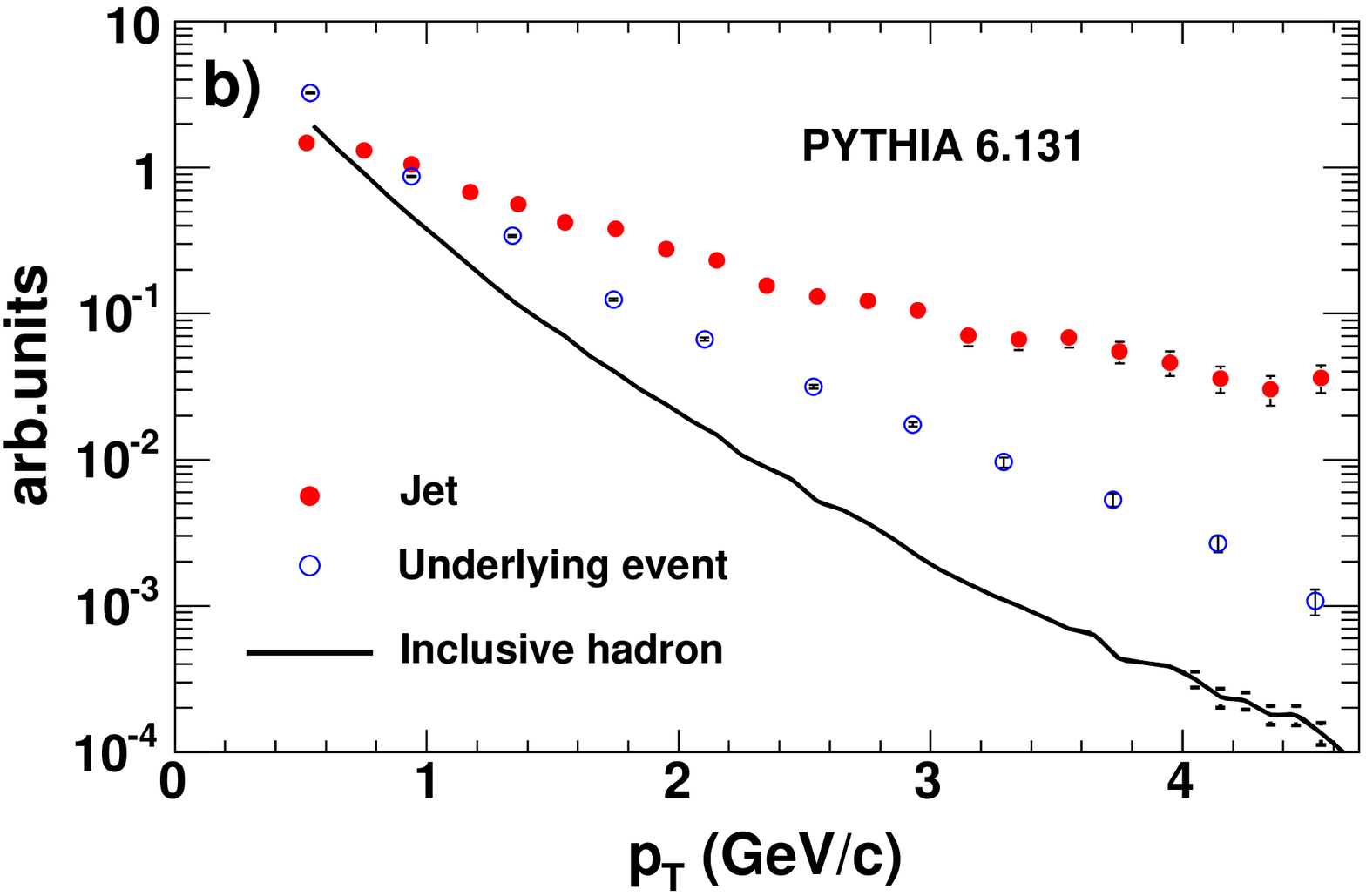,width=0.47\linewidth}
\caption{\label{fig:un2} PYTHIA MC simulation: a)$\Delta\phi$
distribution for $3<p_{T,assoc}<5$ with (top histogram) and
without (bottom histogram) radiation. b) the $p_T$ spectra for jet
pairs and underlying event in $p+p$ events with a $>5$ GeV/$c$
trigger, compared with minimum bias event yield.}
\end{center}
\end{figure}

The underlying event at RHIC can be checked in QCD Monte-carlo
models. We use the PYTHIA6.131, which seems to be able to
reproduce the jet conditional yield as shown in
Figure.\ref{fig:un1}. The roles of the initial/final state
radiations are studied by switching them on and off in PYTHIA
simulation. Figure.\ref{fig:un2}a shows a typical
$\pi^{\pm}-h^{\pm}$ azimuthal correlation with (top histogram) and
without (bottom histogram) radiation from the simulation. There is
a significant enhancement in the pedestal level when radiations
are enabled. We can perform a more quantitative comparison by
plotting separately the jet contribution (double gauss component),
the underlying event (the constant component) and the minimum bias
event level as function of $p_T$ in Figure.\ref{fig:un2}b. The
hierarchy of the three contributions can be clearly seen. For
event tagged with a high $p_T$ jet, the spectra for both the jet
fragmentation and the underlying events are much harder than that
from typical minimum bias events. Current statistics from $p+p$
does not allow a quantitative comparison with the models yet, a
much larger dataset collected from recent RHIC $p+p$ run in 2005
should help to address this question in the near future.

What about the underlying event in $d$ + Au? Figure.\ref{fig:dphi}
indicates that the underlying event levels are larger than those
in $p+p$, although the properties of the jets are quite similar.
Since the hard-scattering only happens in one of the
nucleon-nucleon collision in $d$ + Au, we can assume the ambient
particle production mechanism is the same as in minimum bias
nucleon-nucleon collision. In this case, the ambient particle
production should simply scale as the nuclear modification factor,
$R_{\rm{dAu}}$ measured in $d$ + Au~\cite{Matathias:2005zd}. Thus
the underlying event yields in $p+p$ and $d$ + Au, $U_{\rm{dAu}}$
and $U_{\rm{pp}}$ are connected to each other through the
following simple relation,
\begin{eqnarray}
U_{\rm{dAu}} = U_{\rm{pp}} +
R_{\rm{dAu}}\left(N_{\rm{coll}}-1\right)Yield_{\rm{pp}}
\end{eqnarray}
where Yield$_{\rm{pp}}$ represents the hadron yield per event in
minimum bias $p+p$ collisions. Divide both side by
$\rm{Yield_{pp}}$, we get,
\begin{eqnarray}
\lambda_{\rm{dAu}} = \lambda_{\rm{pp}} + R_{\rm{dAu}}\left(N_{\rm{coll}}-1\right)\\
\lambda_{\rm{dAu}}= U_{\rm{dAu}}/\rm{Yield_{pp}},
\lambda_{\rm{pp}} = U_{\rm{pp}}/\rm{Yield_{pp}}
\end{eqnarray}
note $\lambda_{\rm{pp}}$ denotes the ratio of underlying event
yield to minimum bias event in $p+p$, which should be larger than
1 according to Figure.\ref{fig:dphipp} and Figure.\ref{fig:un2}.

\section{Conclusion}
The di-jet decay kinematics are studied using $\pi^{\pm}-h^{\pm}$
correlation in $p+p$ and $d$ + Au collisions. Measured jet widths,
the calculated $k_T$ and distributions of $p_{out}$ are very
similar between $p+p$ and $d$ + Au, which indicate no or small
broadening in cold nuclear medium. Jet yield distribution in
associated hadron $p_T$ and $x_E$ are also similar between $p+p$
and $d$ + Au, consistent with no significant increase in jet
multiplicity in $d$ + Au relative to $p+p$. The dependence of the
$x_E$ distribution on trigger $p_T$ is weak in measured trigger
$p_T$ range. The underlying event yield in $p+p$ is studied in
PYTHIA Monte-carlo. Events containing a large $p_T$ trigger appear
to have a underlying event spectra much harder than the minimum
bias hadron spectra.

\def\newblock{\hskip .11em plus .33em minus .07em}


\begin{thebibliography}{99}
\expandafter\ifx\csname natexlab\endcsname\relax\def\natexlab#1{#1}\fi
\expandafter\ifx\csname bibnamefont\endcsname\relax
  \def\bibnamefont#1{#1}\fi
\expandafter\ifx\csname bibfnamefont\endcsname\relax
  \def\bibfnamefont#1{#1}\fi
\expandafter\ifx\csname citenamefont\endcsname\relax
  \def\citenamefont#1{#1}\fi
\expandafter\ifx\csname url\endcsname\relax
  \def\url#1{\texttt{#1}}\fi
\expandafter\ifx\csname urlprefix\endcsname\relax\def\urlprefix{URL }\fi
\providecommand{\bibinfo}[2]{#2}
\providecommand{\eprint}[2][]{\url{#2}}

\bibitem{Rak:2004gk}
Rak J \bibinfo{journal}{J.
  Phys.} \textbf{\bibinfo{volume}{G30}}, \bibinfo{pages}{S1309}
  (\bibinfo{year}{2004}), \eprint{hep-ex/0403038}.

\bibitem{Jia:2004sw} Jia J \bibinfo{journal}{J.
  Phys.} \textbf{\bibinfo{volume}{G31}}, \bibinfo{pages}{S521}
  (\bibinfo{year}{2005}), \eprint{nucl-ex/0409024}.

\bibitem{Adler:2002tq}
Adler C \bibnamefont{et~al}
  (\bibinfo{collaboration}{STAR}), \bibinfo{journal}{Phys. Rev. Lett.}
  \textbf{\bibinfo{volume}{90}}, \bibinfo{pages}{082302}
  (\bibinfo{year}{2003}), \eprint{nucl-ex/0210033}.

\bibitem{dan}
Magestro D These proceedings.
\bibitem{wolf}
Holtzmann W These proceedings.

\bibitem{Agakichiev:2003gg}
Agakichiev G \bibnamefont{et~al} (\bibinfo{collaboration}{CERES/NA45}),
  \bibinfo{journal}{Phys. Rev. Lett.} \textbf{\bibinfo{volume}{92}},
  \bibinfo{pages}{032301} (\bibinfo{year}{2004}), \eprint{nucl-ex/0303014}.

\bibitem{Aizawa:2003zq}Aizawa M \bibnamefont{et~al}
  (\bibinfo{collaboration}{PHENIX}), \bibinfo{journal}{Nucl. Instrum. Meth.}
  \textbf{\bibinfo{volume}{A499}}, \bibinfo{pages}{508} (\bibinfo{year}{2003}).

\bibitem{Adler:2003au} Adler S S \bibnamefont{et~al}
  (\bibinfo{collaboration}{PHENIX}), \bibinfo{journal}{Phys. Rev.}
  \textbf{\bibinfo{volume}{C69}}, \bibinfo{pages}{034910}
  (\bibinfo{year}{2004}), \eprint{nucl-ex/0308006}.

\bibitem{Wang:1998ww}
Wang X-N \bibinfo{journal}{Phys. Rev.} \textbf{\bibinfo{volume}{C61}},
  \bibinfo{pages}{064910} (\bibinfo{year}{2000});
Barnafoldi G G, Levai P, Papp G, Fai G I and Zhang Y
  \bibinfo{journal}{Heavy Ion Phys.} \textbf{\bibinfo{volume}{18}},
  \bibinfo{pages}{79} (\bibinfo{year}{2003});
Accardi A\eprint{hep-ph/0312320};
Qiu J-w and Vitev I\bibinfo{journal}{Phys. Lett.} \textbf{\bibinfo{volume}{B570}},
  \bibinfo{pages}{161} (\bibinfo{year}{2003}).

\bibitem{Boer:2003tx}
Boer D and Vogelsang W
  \bibinfo{journal}{Phys. Rev.} \textbf{\bibinfo{volume}{D69}},
  \bibinfo{pages}{094025} (\bibinfo{year}{2004}), \eprint{hep-ph/0312320}.

\bibitem{Hwa:2004sw}
Hwa R C and Yang C B
  \bibinfo{journal}{Phys. Rev.} \textbf{\bibinfo{volume}{C70}},
  \bibinfo{pages}{054902} (\bibinfo{year}{2004}), \eprint{nucl-th/0407081}.

\bibitem{Affolder:2001xt}
Affolder T  et~al (\bibinfo{collaboration}{CDF}), \bibinfo{journal}{Phys. Rev.}
  \textbf{\bibinfo{volume}{D65}}, \bibinfo{pages}{092002}
  (\bibinfo{year}{2002}).

\bibitem{Sjostrand:2004pf} Sjostrand T and Skands P Z
  \bibinfo{journal}{JHEP} \textbf{\bibinfo{volume}{03}}, \bibinfo{pages}{053}
  (\bibinfo{year}{2004}), \eprint{hep-ph/0402078}.

\bibitem{Acosta:2004wq}
Acosta D et al (\bibinfo{collaboration}{CDF}), \bibinfo{journal}{Phys. Rev.}
  \textbf{\bibinfo{volume}{D70}}, \bibinfo{pages}{072002}
  (\bibinfo{year}{2004}), \eprint{hep-ex/0404004}.

\bibitem{Matathias:2005zd}
Matathias F  (\bibinfo{collaboration}{PHENIX}) (\bibinfo{year}{2005}),
  \eprint{nucl-ex/0504019}.

\end{thebibliography}
\end{document}